\documentclass[]{qia_tex/qia}

\graphicspath{{DeComp2/figures/}}

\abstract{Quantum compilers optimize execution-only proxies such as gate count, depth, and fidelity, treating the compiled circuit as the unit of cost. This conflates two distinct resources, how much the substrate has to do at run time, and how much has to be said to describe what to do. Unrolling a looped program leaves run-time cost unchanged while erasing the hierarchical structure on which downstream optimization and reuse rely.

We propose lifting the compiler objective from execution to life cycle complexity by pairing the circuit complexity $\mathcal{C}_{\mathrm{circ}}$ with a Kolmogorov-style description complexity $\mathcal{C}_{\mathrm{Kol}}$ of the emitted circuit, and minimizing the additive total $\mathcal{C}_{\mathrm{tot}} = \alpha\,\mathcal{C}_{\mathrm{circ}} + \beta\,\mathcal{C}_{\mathrm{Kol}}$. The mirrors the additive positional and kinetic entropy of the second law of quantum complexity, and reads as a minimum-description-length regularizer over the otherwise degenerate set of $\mathcal{C}_{\mathrm{circ}}$-minimizers.

We instantiate the objective on $SU(2)$ through exhaustive \texttt{HT}-enumeration with compression-based surrogates upper-bounding $\mathcal{C}_{\mathrm{Kol}}$. Across a Haar-representative target grid the two surrogates correlate positively with $\mathcal{C}_{\mathrm{circ}}$ yet break its rank order on a non-trivial fraction of points, ruling out a tight functional dependence. With calibrated weights the joint cost selects, for a small but operationally meaningful fraction of targets, a candidate that is neither the shortest nor the most compressible \texttt{HT}-string in the $\varepsilon$-ball, exhibiting the compilation choices a $\mathcal{C}_{\mathrm{circ}}$-only optimizer discards and motivating $\mathcal{C}_{\mathrm{Kol}}$ as an active optimization signal for hierarchical intermediate representation for compiler.
}

\begin{document}

\title{\texttt{DeComp}$^\mathtt{2}$: \sffamily Description Complexity aware Decomposition}

\author[]{Aritra Sarkar}

\affiliation[]{Quantum Intelligence Alliance, Kolkata, India}

\date{\today}
\correspondence{Aritra Sarkar at \url{Aritra.Sarkar@quantum-intelligence.net}}
\sourcecode{https://github.com/Advanced-Research-Centre/DeComp2}

\maketitle
\vspace{1em}
\tableofcontents

\newpage
\newcommand{\tocite}[1]{\textcolor{red}{[#1]}}

\section{Introduction}

Quantum compilers optimize executable resources such as gate counts, depth, and fidelity.
This treatment of the compiled circuit's complexity $\mathcal{C}_{\mathrm{circ}}$ as the unit of cost is an incomplete resource ledger. 
A realization of the unitary include both the compiled circuit as well as the program that emits it. 
Consider two circuits that implement the same unitary: an unrolled sequence \texttt{HTHT$^\dagger$HTHT$^\dagger$HTHT$^\dagger$HTHT$^\dagger$} and a one-line program \texttt{repeat\;4:\;HTHT$^\dagger$}. 
Both have identical gate count, depth, non-Clifford count, and end-to-end fidelity, but as computational artifacts, they are not equivalent. 
The looped form admits hierarchy, thereby, composes with other macros, factors through optimization passes, can be reused as a subroutine, and can be verified once and instantiated many times.
The unrolled gate stream loses these information requiring every downstream task to rediscover the structure for efficient reduction in life cycle costs.
$\arg\min \mathcal{C}_{\mathrm{circ}}$ is highly degenerate, with many equally short decompositions of a given $U$, thus $\mathcal{C}_{\mathrm{circ}}$-objectives lack a principled way to prioritize and arbitrate them. 

This article proposes lifting the compiler's objective from execution complexity to life cycle complexity by pairing $\mathcal{C}_{\mathrm{circ}}$ with a description-length term $\mathcal{C}_{\mathrm{Kol}}$ on the program that emits the gate sequence. 
Thus, it can be positioned as the minimum-description-length principle \cite{rissanen1978modeling} being applied to quantum compilation.
A good decomposition balances how much the substrate has to do against how much has to be said to describe what to do.
This trade-off drives tasks such as grammar inference, dictionary learning, symbolic regression, and program synthesis. 
In an additive form, the total complexity, $\mathcal{C}_{\mathrm{tot}} = \alpha\,\mathcal{C}_{\mathrm{circ}} + \beta\,\mathcal{C}_{\mathrm{Kol}}$, include the description term acting as a secondary regularizer over the otherwise degenerate set of $\mathcal{C}_{\mathrm{circ}}$-minimizers.
This regularizer is an Occam's-razor penalty, analogous in spirit to an $L_1$ regularizer in statistical learning, that prefers structured, short-to-describe decompositions among the equally short ones. 

This engineering-motivated additive ledger echoes the structure of the second law of quantum complexity \cite{brown2018second}, which decomposes the entropy of an auxiliary classical system into a positional and a kinetic part.
That conjecture posits that the computational complexity of a generic quantum state grows linearly in time until it saturates near $\sim \exp(S)$, then fluctuates and eventually recurs, in analogy with the thermodynamic second law for coarse-grained entropy. 
The auxiliary system carries $2^K$ degrees of freedom whose phase-space entropy decomposes as $S_{\mathrm{tot}} = S_{\mathrm{pos}} + S_{\mathrm{kin}}$, with the positional entropy corresponding to the quantum circuit complexity $\mathcal{C}_{\mathrm{circ}}(U)$ and the kinetic entropy to the description complexity $K(H)$ of the generating Hamiltonian.
We borrow only this additive decomposition of complexity into execution and description components as a design principle, however, do not rely on the associated monotonic complexity-growth conjecture.
Separately, the Complexity = Action (CA) \cite{brown2016holographic} and Complexity = Volume (CV) \cite{susskind2016computational} holographic dualities identify the single boundary quantity $\mathcal{C}_{\mathrm{circ}}(U)$ with a single bulk geometric object (the on-shell gravitational action of the Wheeler-DeWitt patch, and the maximal-volume slice, respectively), with no description-length companion on the bulk side.

The proposal of this article is to enrich quantum complexity to an engineering artifact, by designing a quantum unitary decomposition procedure whose optimization target combines a circuit term and a description term. 
In particular, in addition to the usual gate count, depth, and fidelity, we co-optimize for the description term defined as the Kolmogorov complexity $\mathcal{C}_{\mathrm{Kol}}(p)$ of the quantum program that emits the gate sequence $C = g_1\cdots g_m$ drawn from a discrete gate set $\mathcal{G}$ within an approximation tolerance $\varepsilon$ for the target unitary, rather than $K(H)$ for a continuous $U \in SU(2)$. 
An open-source implementation of the resulting $\mathcal{C}_{\mathrm{tot}}$-guided decomposer accompanies this article and is directly usable in a quantum-compiler pipeline.

The rest of this article is organized as follows.
\S\ref{sec:bulk} develops the total-complexity formalism, casting $\mathcal{C}_{\mathrm{circ}}$ as a discrete positional-entropy proxy and $\mathcal{C}_{\mathrm{Kol}}$ as its kinetic-entropy companion, formulating the additive objective of Equation~\eqref{eq:action}, and situating it among related joint-complexity measures.
\S\ref{sec:exp} instantiates the objective on a deterministic $SU(2)$ testbed, establishing the independence of the two cost axes and exhibiting targets whose joint minimizer differs from both per-axis minimizers.
\S\ref{sec:future} outlines lifting $\mathcal{C}_{\mathrm{Kol}}$ from a passive surrogate to an active signal that drives a hierarchical compiler IR.
\S\ref{sec:conclusion} concludes the article.

\section{Total complexity formalism}
\label{sec:bulk}

In this section, we revisit the background and formalize $\mathcal{C}_{\mathrm{tot}}$ guided unitary decomposition in the pragmatic discrete setting.
First, we cast $\mathcal{C}_{\mathrm{circ}}$ as the discrete proxy for positional entropy, with Nielsen-geometric, space-time pebbling, and quantum speed limit consideration.
Then, kinetic-entropy half in introduced and substituted with $\mathcal{C}_{\mathrm{Kol}}(p)$ for the compiler setting.
The compiler objective is formulated and its relation to other joint complexity measures is discussed.

\subsection{Circuit complexity $\leftrightarrow$ positional entropy}
\label{sec:circ}

The \textit{Complexity $\leftrightarrow$ Action} conjecture \cite{brown2016holographic} pairs the boundary computational complexity of a quantum state with the on-shell gravitational action of the Wheeler-DeWitt patch in the bulk,
\begin{equation}
  \mathcal{C}_{\mathrm{circ}}(U\ket{\psi_0}) \;\stackrel{?}{=}\; \frac{1}{\pi\hbar}\,I_{\mathrm{WDW}}.
\end{equation}
This structural duality relates the circuit complexity observable that grows monotonically on the boundary with a geometric integral (an action) that accumulates along a bulk patch.

For our discrete formulation, we define,
\begin{equation}
  \mathcal{C}_{\mathrm{circ}}(U) \;=\; \min_{\,g_1 \cdots g_m\,} \sum_{i=1}^{m} \ell(g_i), \qquad g_i \in \mathcal{G}, \qquad \|g_1\cdots g_m - U\| \leq \varepsilon,
  \label{C_circ}
\end{equation}
i.e., the minimal weighted gate length realizing $U$ to within operator-norm (or final-state) tolerance $\varepsilon$.
The weight $\ell(g)$ may itself encode locality structure, inheriting the Nielsen penalty tensor in the continuum picture.
In the holographic correspondence this weighted gate count is the discrete proxy for the positional entropy $S_{\mathrm{pos}}$, measuring how much of the unitary-group configuration space the synthesis sequence had to traverse to reach $U$.
Operationally, it quantifies the space-time volume (depth for fixed space/qubit budget) of work the quantum substrate does per execution.
For a unitary assembled from Haar-random two-qubit gates, \cite{haferkamp2022linear} prove that the exact circuit complexity grows linearly in the number of gates, with unit probability, until it saturates after exponentially many gates, thereby settling the Brown-Susskind conjecture \cite{brown2018second} on the positional-entropy axis.
The description term $\mathcal{C}_{\mathrm{Kol}}$ however remains an independent, uncontrolled axis.

With a discrete $\mathcal{G}$ the Solovay-Kitaev theorem \cite{dawson2005solovay} guarantees existence of such a sequence with $m=O(\log^{c}(1/\varepsilon))$ for any fixed $U$.
We note three further considerations in estimating $\mathcal{C}_{\mathrm{circ}}(U)$.

\subsubsection*{Nielsen's geodesic}

In the continuum limit $\mathcal{C}_{\mathrm{circ}}$ becomes the geodesic length \cite{nielsen2005geometric,nielsen2006quantum,nielsen2006optimal} in a right-invariant Finsler metric whose penalty tensor suppresses generators acting on more than two qubits,
\begin{equation}
  d(I,U) \;=\; \inf_{H(t)} \int_{0}^{T}\! \sqrt{\langle H(t),H(t)\rangle_{G}}\;dt, \qquad U(T)=U,
\end{equation}
with the inner product $\langle H,H\rangle_G = \sum_{I,J} G_{IJ}\,h^I h^J$ defined by expanding the time-dependent generator on a basis of $\mathfrak{su}(2^n)$, $H(t) = \sum_I h^I(t)\,\sigma_I$, where the indices $I,J$ run over the $4^n-1$ non-identity Pauli strings. In Nielsen's penalty-tensor convention $G_{IJ}$ is diagonal, with $G_{II}=1$ for Pauli weight $\le 2$ and $G_{II}=q\gg 1$ for weight $>2$, so the metric punishes motion along $k$-local directions with $k>2$.

We use this picture to motivate the structure of $\ell(g)$, i.e., the per-primitive weight should mirror the locality structure of $G_{IJ}$, i.e., cheap for $1$- and $2$-local gates and expensive for $k>2$ primitives. 
Note that, in our case, the compiler searches over discrete sequences in $\mathcal{G}$ with the bound $\varepsilon$ on $\|g_1\cdots g_m - U\|$ relaxing the continuum requirement of $U(T)=U$.

Quantum optimal control solves the $\mathcal{C}_{\mathrm{circ}}$-only problem through mature open- and closed-loop pulse-engineering methods, including GRAPE \cite{khaneja2005optimal}, CRAB \cite{caneva2011chopped}, Krotov method \cite{fonseca2022effectiveness}, GOAT \cite{machnes2018tunable}, and RL-controllers \cite{niu2019universal}.
These methods minimize a geodesic-like fidelity-plus-gate-time functional tied to the Nielsen geometry.
Energy-optimal gradient ascent pulse engineering (EO-GRAPE) \cite{fauquenot2025open} augments the GRAPE update with a gradient of the pulse energetic cost, defined as the time-integrated norm of the control Hamiltonian, tracing a Pareto front between energetic cost and fidelity.
The same work shows that, for a single-qubit gate, this energetic cost correlates positively with the length of the path traced on the Bloch sphere, so the most energy-efficient pulses approximate the geodesic between the initial and target states, with structured EO-GRAPE pulses following near-geodesic arcs.

\subsubsection*{Bennett's space-time pebbling}
\label{sec:bennett}

Gate count is a space-restrictive estimate of circuit complexity. 
We promote the discrete circuit cost to a spatio-temporal variant that accounts for ancilla and idle qubits as well as gate primitives,
\begin{equation}
  \mathcal{C}_{\mathrm{circ}}^{\,\mathrm{ST}}(U) \;=\; \min_{C\rightsquigarrow_\varepsilon U} \Bigl[\,\text{depth}(C) * \text{width}(C)\,\Bigr],
\end{equation}
where the minimization is over circuits $C$ over $\mathcal{G}$ that approximate $U$ within $\varepsilon$. 
Reversible/quantum compilation must uncompute ancillae, and the space-time trade-off is exemplified by Bennett's pebble game on the computation DAG \cite{bennett1989time,meuli2019reversible}.
$k$ pebbles on an $n$-step computation can simulate it in time $O(n^{1+\epsilon(k)})$ with $\epsilon(k)\to 0$ as $k\to\infty$. 
Plugging this relation into $\mathcal{C}_{\mathrm{circ}}^{\,\mathrm{ST}}$ recovers known ancilla/depth trade-offs as the Pareto front.

\subsubsection*{Quantum speed limits}

The discrete schedule cannot run arbitrarily fast.
Across any fixed energy budget, the Mandelstam-Tamm \cite{mandelstam1991uncertainty} and Margolus-Levitin \cite{margolus1998maximum} bounds impose,
\begin{equation}
  T_{\min} \;\geq\; \frac{\pi\hbar}{2}\, \max\!\left(\frac{1}{\Delta E},\;\frac{1}{\langle E\rangle - E_{0}}\right),
\end{equation}
which translates into a lower bound on the wall-clock time of a circuit of given depth and per-gate energy. 

The QSL acts as a global feasibility filter on candidate schedules \cite{pires2016generalized,deffner2017quantum}.
Any sequence that violates the QSL for the device's accessible energy scale can be dropped from the search. 

\subsection{Description complexity $\leftrightarrow$ kinetic entropy}
\label{sec:descr}

Brown-Susskind \cite{brown2018second} identify the kinetic-entropy half of the auxiliary classical system with the Kolmogorov complexity of the generating Hamiltonian,
\begin{equation}
  K(H) \;=\; \min \,\big\{\,|q| \;:\; \mathcal{U}(q) \text{ outputs (an encoding of) } H\,\big\},
\end{equation}
on a fixed universal classical reference machine $\mathcal{U}$ \cite{li2008introduction}. 
In the auxiliary-system picture, $K(H)$ is the spread of the classical degree of freedom along the momentum directions, i.e., a more complex generator excites more momentum components, raising the kinetic entropy. 

A Hamiltonian with a short Pauli expansion (e.g., a $k$-local model with $O(\mathrm{poly}(n))$ coefficients given by a closed formula) has small $K(H)$; a generic dense Hermitian operator has $K(H) = \Omega(4^n)$. 
A basis-independent count measures the dimension of the dynamical Lie algebra $\mathfrak{g} = \langle iH_1,\dots,iH_L\rangle_{\mathrm{Lie}}$ generated by the terms of $H$ under commutation \cite{zeier2011symmetry,wiersema2023classification}, or the Krylov subspace dimension spanned by the nested commutators $\{H,\,[H,O],\,[H,[H,O]],\dots\}$ \cite{parker2019universal,rabinovici2021operator}.
When $\dim\mathfrak{g} = O(\mathrm{poly}(n))$, the induced evolution $e^{-iHt}$ is confined to a polynomial-dimensional sub-manifold and admits a compact Lie-algebraic representation \cite{somma2006efficient,goh2025lie}, and the Krylov dimension is the operator-space realization of the momentum spread.
However, these dimensions measure the complexity of the dynamics generated by $H$ rather than the literal description length of $H$ itself, since a few-term local Hamiltonian with small $K(H)$ can still generate the full $\mathfrak{su}(2^n)$, so the dynamical Lie algebra and Krylov dimensions bound the cost of representing the induced evolution and complement, rather than equal, the Pauli-expansion length.

For a discrete compiler the generator $H$ is not a natural representation.
We instead consider decomposition algorithms $p$ (for example, Solovay-Kitaev \cite{dawson2005solovay} or Ross-Selinger \cite{ross2016optimal}) that emits a discrete gate sequence, and take the description cost to be the Kolmogorov complexity of the emitted circuit,
\begin{equation}
  \mathcal{C}_{\mathrm{Kol}}(U) \;=\; \min \,\big\{\,|p| \;:\; \mathcal{U}(p) \mapsto g_1\cdots g_m \,\big\}, \qquad g_i \in \mathcal{G}, \qquad \|g_1\cdots g_m - U\| \le \varepsilon,
  \label{C_Kol}
\end{equation}
i.e., the length of the shortest program on the reference machine $\mathcal{U}$ whose output is the gate sequence $g_1\cdots g_m$, subject to realizing $U$ in operator norm to tolerance $\varepsilon$.

It is important to note that, while $K(H)$ and $\mathcal{C}_{\mathrm{Kol}}(U)$ both quantify descriptions of a generator of the dynamics on a fixed universal machine, the latter allow an independent degree of freedom.
$K(H)$ describes the continuous physical Hamiltonian of $U$ and thus uniquely maps to it.
$\mathcal{C}_{\mathrm{Kol}}(U)$, in contrast, describes the discrete classical controller and can be influenced by choosing a compilation strategy. 
Two different decompositions of the same $U$ both within the approximation bound and similar gate resources (i.e., $\mathcal{C}_{\mathrm{circ}}(U)$), can have vastly different program (or compressed) representation.

The energy-efficient quantum instruction set architecture (EQISA) of \cite{mishra2026eqisa} is among the few works to consider description complexity term into a quantum computing stack.
It uses compression (\texttt{bzip2}) as a tractable proxy for Kolmogorov complexity, paired with Huffman encoding over a sparse learned dictionary of recurring gate fragments, and independently evaluates the circuit and the description complexities for the MQT Bench suite of algorithms \cite{quetschlich2023mqt}.

\subsection{Compiler objective}

The second-law paper \cite{brown2018second} introduces a holographic dual classical system with $2^K$ degrees of freedom whose phase-space entropy tracks the second law of complexity. 
Its total entropy splits into two additive orthogonal terms in phase-space,
\begin{equation}
  S_{\mathrm{tot}} \;=\; \underbrace{S_{\mathrm{pos}}}_{\,\leftrightarrow\;\mathcal{C}_{\mathrm{circ}}(U)\,} \;+\; \underbrace{S_{\mathrm{kin}}}_{\,\leftrightarrow\;K(H)\,}
\end{equation}

We adopt this additive form, with the $ \mathcal{C}_{\mathrm{Kol}}(p)$ substitution explained above.
Thus, we define total complexity,
\begin{equation}
  \mathcal{C}_{\mathrm{tot}}(U) \;=\; \alpha\,\mathcal{C}_{\mathrm{circ}}(U) \;+\; \beta\,\mathcal{C}_{\mathrm{Kol}}(U)
  \label{eq:totalC}
\end{equation}
as the figure of merit for compilation.
The two orthogonal and additively resources involve:
\begin{itemize}[nolistsep,noitemsep]
    \item $\mathcal{C}_{\mathrm{circ}}$ is a count on the unitary group (run-time on the quantum substrate) 
    \item $\mathcal{C}_{\mathrm{Kol}}$ is a description length on a classical reference machine (storage and streaming cost on the controller) 
\end{itemize} 

Substituting Equations~\ref{C_circ} and \ref{C_Kol} in Equation \ref{eq:totalC},
\begin{equation}
  \boxed{\;\mathcal{C}_{\mathrm{tot}}(U,p) \;=\; \alpha\min_{\,g_1 \cdots g_m\,} \sum_{i=1}^{m} \ell(g_i) \;+\; \beta\;\min \,\big\{\,|p| \;:\; \mathcal{U}(p) \mapsto g_1\cdots g_m \,\big\}, \qquad g_i \in \mathcal{G}, \quad \|g_1\cdots g_m - U\| \le \varepsilon\;}
  \label{eq:action}
\end{equation} 
The gate sequence that minimizes $\mathcal{C}_{\mathrm{circ}}$ and the one that minimizes $\mathcal{C}_{\mathrm{Kol}}$ within the $\varepsilon$ tolerance are in general distinct, and both may differ from the sequence that minimizes the joint cost $\mathcal{C}_{\mathrm{tot}}$.
This separation between the per-axis optima and the joint optimum is the design freedom that our proposed compiler exploits.
Note, $\alpha=\beta=1$ is considered throughout this note, deferring weight calibration to a deployment-level concern.

Taken together, EO-GRAPE \cite{fauquenot2025open} and EQISA \cite{mishra2026eqisa} anchor the abstract resource counts of \eqref{eq:action} to physical cost on either side of the quantum-classical interface.
EO-GRAPE ties $\mathcal{C}_{\mathrm{circ}}$ to the energetic cost of the control pulses on the quantum substrate, while EQISA ties $\mathcal{C}_{\mathrm{Kol}}$ to the storage and streaming energy cost of the instructions from the classical controller.
The $\mathcal{C}_{\mathrm{tot}}$ objective thereby directs a total energetic cost optimizer.

\subsection{Related formalisms}
\label{sec:joint}

This subsection situates total complexity within related measures that consider a description term alongside an execution term.

\subsubsection*{Levin complexity}
Levin's $Kt(x) = \min_{p}\bigl(|p| + \log t(p)\bigr)$ \cite{levin1973universal} is the closest classical analogue, additively balancing description length against (log) runtime to drive universal search.
Identifying $|p|\equiv \mathcal{C}_{\mathrm{Kol}}$ and $\log t(p) \equiv \log \mathcal{C}_{\mathrm{circ}}^{\,\mathrm{ST}}$ recovers the additive structure of Equation \eqref{eq:action}, with a logarithmic rather than linear weight for execution cost.
The upshot of $\mathcal{C}_{\mathrm{tot}}$ is that, it is not an ad-hoc linear combination but the quantum-compiler instance motivated by holographic arguments from quantum cosmology.

\subsubsection*{Schmidhuber's speed prior}

The speed prior $S(x) \propto \sum_{p:\mathcal{U}(p)=x} 2^{-|p|}/t(p)$ \cite{schmidhuber2002speed} biases inference toward hypotheses that are simultaneously short and fast.
Used as a regulariser, $-\log S(p)$ reproduces the additive structure of $\mathcal{C}_{\mathrm{tot}}$, with $\alpha=\beta\ln 2$, so compilation can be read as maximum-a-posteriori inference under a speed prior over programs.

\subsubsection*{Quantum circuit probability}

Solomonoff's algorithmic probability \cite{solomonoff1964formal} of a string equates to $2^{-\mathcal{C}_{\mathrm{Kol}}}$, tying the description to a universal prior over programs.
The coding theorem \cite{levin1974laws} allows empirically estimating it.
Visualizing quantum circuit probabilities \cite{bach2023visualizing} carries this correspondence to the circuit setting by weighting gate sequences by their occurrence probability under a random-circuit ensemble, so that frequent, compressible circuits carry low description cost.
Reading $\mathcal{C}_{\mathrm{Kol}}$ as a negative log-probability recovers the additive $\mathcal{C}_{\mathrm{tot}}$ as a maximum-a-posteriori objective, consistent with the speed-prior view above.

\subsubsection*{Baez's algorithmic thermodynamics}

Algorithmic thermodynamics \cite{baez2012algorithmic,ebtekar2025foundations} places a Gibbs ensemble over halting programs, so that a program's description length plays the role of energy and Kolmogorov complexity becomes a free energy.
The partition function $Z = \sum_p 2^{-|p|}$ is Chaitin's halting probability, and the resulting canonical ensemble assigns each program a Boltzmann weight in its length.
Reading $\mathcal{C}_{\mathrm{Kol}}$ as this energy makes the weights $\alpha$ and $\beta$ conjugate inverse temperatures that set the exchange rate between description and execution.
In this perspective the calibration of $\beta$ in our experiments is a choice of temperature, and the low-temperature limit recovers the description-dominated regime in which the joint cost collapses onto $\arg\min\,\mathcal{C}_{\mathrm{Kol}}$.

\subsubsection*{Toffoli's Lagrangian formulation and Selesnick's Yang-Mills action for quantum circuits}

A complementary least-action view counts computation by an action rather than an entropy.
The Lagrangian formulation \cite{toffoli2003lagrangian} and Yang-Mills action for quantum circuits \cite{selesnick2014action} both assign an action integral to a gate sequence, recovering the space-time cost $\mathcal{C}_{\mathrm{circ}}^{\,\mathrm{ST}}$ as the on-circuit analogue of the Complexity $\leftrightarrow$ Action reading of $\mathcal{C}_{\mathrm{circ}}$.
Unlike the entropy-based measures above, these formulations refine only the execution term, promoting the flat gate count to a locality-weighted action over the circuit worldsheet.
They therefore supply a principled origin for the weight $\ell(g)$ of Equation~\eqref{C_circ} while leaving the description axis $\mathcal{C}_{\mathrm{Kol}}$ untouched, and are complementary to, rather than a substitute for, the additive $\mathcal{C}_{\mathrm{tot}}$.

\subsubsection*{Bennett's logical depth}

The logical depth of a string $x$ is the runtime of the shortest program that outputs $x$ \cite{bennett1988logical}, explicitly coupling $\mathcal{C}_{\mathrm{Kol}}$ to compute time.
Unlike the additive $\mathcal{C}_{\mathrm{tot}}$, logical depth aggregates the two axes lexicographically, first minimizing the description and then measuring the runtime of that specific minimiser.
It is therefore recovered as a corner of the additive Pareto front in the limit $\sfrac{\alpha}{\beta}\to 0$.

\subsubsection*{Gell-Mann-Lloyd effective complexity}

Effective complexity \cite{gell1996information,ay2010effective} measures the algorithmic information content of a string's regularities alone, separating structure from incidental randomness through a two-part code that names an ensemble of which the string is typical and charges only the description length of that ensemble.
It refines the description axis and is non-monotone in $\mathcal{C}_{\mathrm{Kol}}$.
An algorithmically random string has near-maximal $\mathcal{C}_{\mathrm{Kol}}$ yet near-zero effective complexity, since it has no regularities to describe \cite{ay2010effective}.

\subsubsection*{Thermodynamic depth and statistical complexity}

Further adjacent measures include thermodynamic depth \cite{lloyd1988complexity} and statistical complexity $C_\mu$ \cite{crutchfield1989inferring}.
Both refine only the execution side, quantifying the history-dependence of reaching $U$ ($\mathcal{C}_{\mathrm{circ}}$-like) without a description-length companion, and so sit furthest from our scope.

\subsubsection*{Resource-bounded universal agents}

The pairing of a description term with an execution term also underlies universal artificial general intelligence. 
The AIXI-tl agent \cite{hutter2000theory} bounds the incomputable AIXI by a program length $l$ and a runtime $t$, ranking policies by description length against a $t\,2^l$ execution budget.
The quantum knowledge-seeking agent QKSA \cite{sarkar2022qksa} ports this to quantum environments, scoring quantum-process-tomography programs by a mutable multi-objective LEAST cost that combines algorithmic complexity with computational-resource complexity, while QAIXI \cite{perrier2025quantum} carries the AIXI value function itself into quantum information.
\cite{sarkar2026tao} argues that the weights in such a joint cost, and the rewards they trade off, are far from canonical.
QKSA's evolution of the cost function by genetic programming towards recursive self-improvement is also pursued by the Darwin Godel machine \cite{zhang2025darwin} and hyperagents \cite{zhang2026hyperagents} in augmenting Schmidhuber's Godel machine \cite{schmidhuber2007godel}.
These suggests that the $(\alpha, \beta)$ in $\mathcal{C}_{\mathrm{tot}}$ could likewise be empirically determined.

\section{Empirical study on $SU(2)$}
\label{sec:exp}

In this section, we instantiate the developed formalism on a tractable single-qubit testbed to exemplify the claim that: for a fixed target $U$ there exist multiple HT-sequences inside the $\varepsilon$-tolerance ball whose individual $\mathcal{C}_{\mathrm{circ}}$ and $\mathcal{C}_{\mathrm{Kol}}$ disagree on the ranking, so that the joint cost $\mathcal{C}_{\mathrm{tot}} = \alpha\mathcal{C}_{\mathrm{circ}} + \beta\mathcal{C}_{\mathrm{Kol}}$ selects a non-trivial decomposition that neither component would have picked independently.
Thus, it exhibits an instance in which a $\mathcal{C}_{\mathrm{circ}}$-only optimiser, of the kind that quantum compilers and quantum optimal control techniques typically provide, returns a suboptimal candidate.

\subsection{Target ensemble on $SU(2)$}
\label{sec:exp-sampling}

A target $U \in SU(2)$ is parameterised by an axis $\hat n \in S^2$ and a rotation angle $\theta \in [0,2\pi]$ via $U = \exp(-i\,\theta\,\hat n\cdot\vec\sigma/2)$. The ensemble draws
\begin{itemize}[nolistsep,noitemsep]
  \item axes $\hat n_k$ from a Fibonacci lattice on $S^2$ (deterministic, near-uniform), indexed $k = 1,\dots,x$;
  \item angles $\theta_j$ from the Haar-induced marginal $p(\theta) \propto \sin^2(\theta/2)$ on $[0,2\pi]$, with $y$ samples per axis generated deterministically by inverse cumulative distribution function (CDF) on equispaced quantiles: writing the normalised CDF $F(\theta) = (\theta - \sin\theta)/(2\pi)$, $\theta_j = F^{-1}\!\bigl((j-\tfrac12)/y\bigr)$ for $j = 1,\dots,y$, with $F^{-1}$ evaluated by Newton iteration on the monotone $F$. The empirical distribution of $\{\theta_j\}$ converges weakly to $p(\theta)$ as $y \to \infty$, with $O(1/y)$ Kolmogorov-Smirnov error rather than the $O(1/\sqrt y)$ of i.i.d. sampling.
\end{itemize}

\begin{figure}[!hbt]
  \centering
  \begin{subfigure}[t]{0.28\linewidth}
    \centering
    \includegraphics[width=\linewidth]{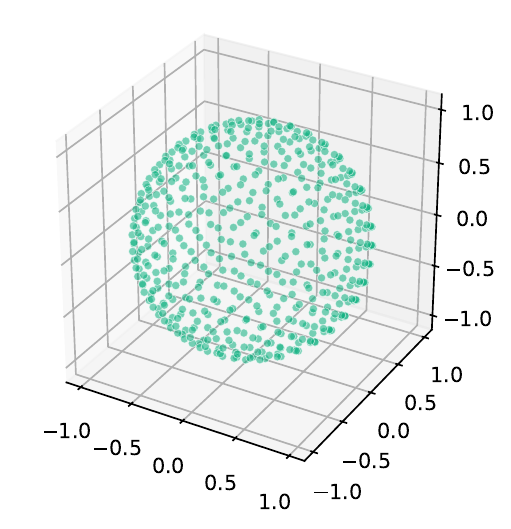}
    \caption{Fibonacci lattice on $S^2$ used as the axis grid $\{\hat n_k\}$.}
    \label{fig:axes}
  \end{subfigure}\hfill
  \begin{subfigure}[t]{0.68\linewidth}
    \centering
    \includegraphics[width=\linewidth]{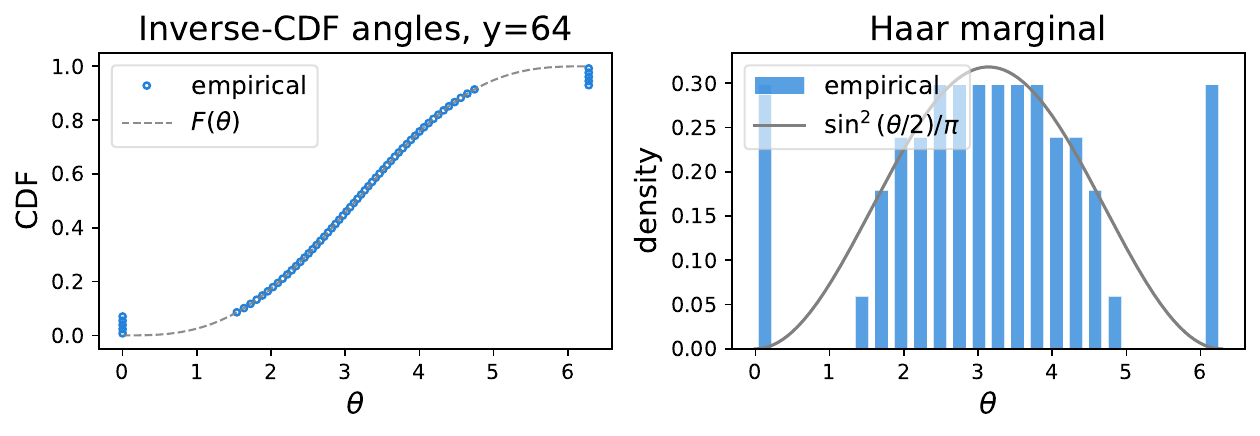}
    \caption{Inverse-CDF angles $\{\theta_j\}$: empirical CDF against $F(\theta) = (\theta-\sin\theta)/(2\pi)$ (left) and density against the Haar marginal $\sin^2(\theta/2)/\pi$ (right).}
    \label{fig:angles}
  \end{subfigure}
  \caption{Deterministic target ensemble of $SU(2)$: axis grid (left) and Haar-induced angle grid (right) at $(x,y)=(512,64)$.}
  \label{fig:ensemble}
\end{figure}

The Fibonacci grid on $\hat n$ gives a reproducible axis, while the $\sin^2(\theta/2)$ weight on $\theta$ maintains a Haar sample on $SU(2)$. 
As a sanity-check, we draw uniform unit quaternions on $S^3$, the exact Haar measure on $SU(2)$, and compares the resulting scatter to the Fibonacci-grid scatter. 
Figure~\ref{fig:ensemble} displays the deterministic ensemble used in the experiments. 
The Fibonacci lattice in Figure~\ref{fig:axes} populates $S^2$ uniformly without clustering at the poles, and the inverse-CDF angles in Figure~\ref{fig:angles} track the Haar marginal $\sin^2(\theta/2)/\pi$ already at modest $y$, with the empirical CDF lying on top of $F(\theta)$.

\subsection{Complexity surrogates}
\label{sec:exp-surrogates}

For each target $U$ on the grid we obtain a $(\widehat{\mathcal{C}}_{\mathrm{circ}},\widehat{\mathcal{C}}_{\mathrm{Kol}})$ pair through two complementary routes: exhaustive HT-enumeration, which is brute-force but optimal up to the enumeration cap, and an algorithmic decomposition baseline, which scales but is suboptimal.

\subsubsection{Exhaustive enumeration}

Binary strings $s \in \{0,1\}^*$ are enumerated in length-first order under the map $0 \mapsto \texttt{H}$, $1 \mapsto \texttt{T}$. For each $s$ the unitary $U_s = g_{s_{|s|}}\cdots g_{s_1}$ is built incrementally and, for every target $U$, the shortest $s$ with $\|U_s - U\| \le \varepsilon$ is recorded together with the full list of in-tolerance alternatives. The exhaustive surrogates are
\begin{equation}
  \mathcal{C}_{\mathrm{circ}}^{(1)}(U) = |s|, \hspace{6em} \mathcal{C}_{\mathrm{Kol}}^{(1)}(U) = \mathrm{compress}(s), \quad \mathrm{compress} \in \{\mathrm{bzip2}, \mathrm{pybdm}\}.
\end{equation}
Enumeration costs $2^{M+1}-2$ unitary multiplications to scan all strings up to length $M$, which fixes the run-time budget.
We diagnose:
\begin{itemize}[nolistsep,noitemsep]
  \item coverage: fraction of the $x\cdot y$ target grid that admits at least one $s$ with $|s| \le M$ and $\|U_s - U\| \le \varepsilon$;
  \item length spectrum: empirical distribution of the realised minimal $|s|$ over the covered targets.
\end{itemize}
These allow us to estimate the smallest $M$ that gives non-trivial coverage at a given $(x,y,\varepsilon)$ while bounding the enumeration cost. 
The presented experiments sets $(x,y,\varepsilon,M)=(512,64,0.15,12)$.
Thus, the enumerator scans $2^{M+1}-2=8190$ \texttt{HT}-strings up to length $M=12$ and retains, for each target on the $x\cdot y$ grid, the list of strings inside the operator-norm ball. 
Coverage on this grid is $20\,327/32\,768 \approx 62\%$.

\subsubsection{Decomposition algorithms}

A second estimate is obtained by decomposing $U$ with Solovay-Kitaev decomposition (SKD) and pygridsynth (Ross-Selinger) to tolerance $\varepsilon$. 
$s_{\mathrm{syn}}$ denotes the resulting \texttt{HT}-string,
\begin{equation}
  \mathcal{C}_{\mathrm{circ}}^{(2)}(U) = |s_{\mathrm{syn}}|, \quad \mathrm{syn} \in \{\mathrm{SKD}, \mathrm{GridSynth}\} \hspace{6em} \mathcal{C}_{\mathrm{Kol}}^{(2)}(U) = \mathrm{compress}(s_{\mathrm{syn}}).
\end{equation}
The synthesised length depends on algorithm hyperparameters (precision level, recursion depth, etc.). The $\alpha$ and $\beta$ weight pair is left unconstrained for diagnostic plots ($\alpha=\beta=1$) and pinned to a calibrated value to control the witness count.

\subsection{$(\mathcal{C}_{\mathrm{circ}},\mathcal{C}_{\mathrm{Kol}})$ independence}
\label{sec:exp-viz}

To capture the three-dimensional structure of $SU(2)$, the experiment presents two complementary 2D views of the same enumerated data: a scatter in the cost plane and heatmaps over the $(k,\theta)$ parameter grid,
concerning different aspects of the claim.

\begin{figure}[htb]
  \centering
  \includegraphics[width=0.65\linewidth]{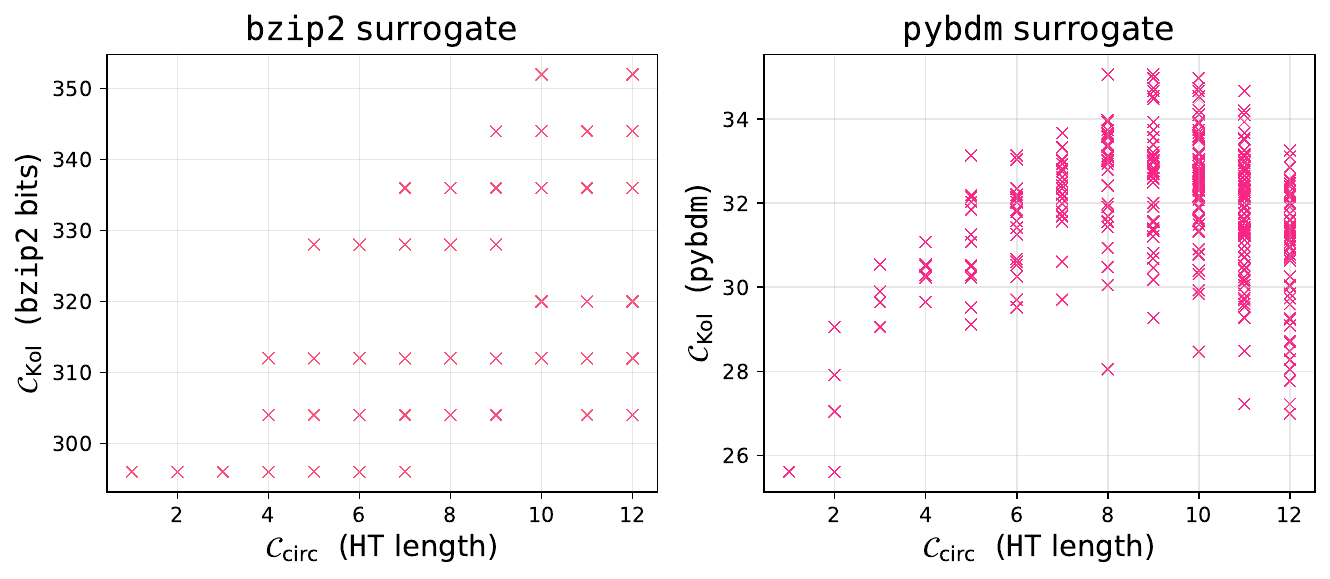}
  \caption{$(\mathcal{C}_{\mathrm{circ}}, \mathcal{C}_{\mathrm{Kol}})$ scatter, one point per Fibonacci-grid target covered by the \texttt{HT} enumeration at $(x,y,\varepsilon,M)=(512,64,0.15,12)$. Left: \texttt{bzip2} bit-length surrogate of $\mathcal{C}_{\mathrm{Kol}}$. Right: \texttt{pybdm} block-decomposition surrogate. Off-diagonal spread under both surrogates supports the independence-of-axes claim~(i).}
  \label{fig:scatter}
\end{figure}

\paragraph{Non-collinearity.} 
We use \texttt{bzip2} and \texttt{pybdm} as the upper-bound estimator of $\mathcal{C}_{\mathrm{Kol}}$ \cite{seward1996bzip2,talaga2024pybdm,soler2014calculating}.
As shown in Figure~\ref{fig:scatter}, both surrogates correlate positively with $\mathcal{C}_{\mathrm{circ}}$ (Pearson $r \approx +0.7$ for \texttt{bzip2}) but place the targets on visibly different vertical scales and break the rank order on a non-trivial fraction of the covered points. 
A tight functional dependence is therefore ruled out, i.e., $\mathcal{C}_{\mathrm{Kol}}$ carries information that $\mathcal{C}_{\mathrm{circ}}$ does not. 

\begin{figure}[htb]
  \centering
  \includegraphics[width=0.95\linewidth]{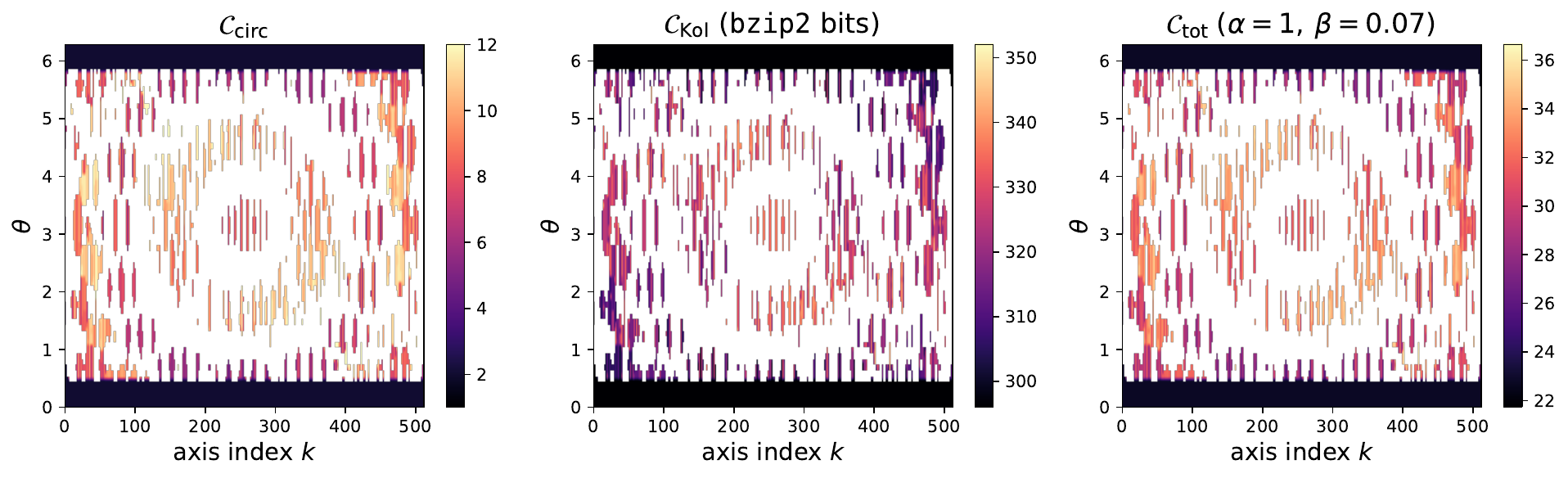}
  \caption{Heatmaps of $\mathcal{C}_{\mathrm{circ}}$, $\mathcal{C}_{\mathrm{Kol}}$ (\texttt{bzip2}) and $\mathcal{C}_{\mathrm{tot}}$ over the axis index $k\in\{1,\dots,x\}$ and rotation angle $\theta\in[0,2\pi]$ at $(x,y,\varepsilon,M)=(512,64,0.15,12)$. The two cost fields are qualitatively different, so $\mathcal{C}_{\mathrm{tot}}$ is not a monotone reshading of either component alone.}
  \label{fig:heatmaps}
\end{figure}

\paragraph{Structural dissimilarity.} 
Plotting $\mathcal{C}_{\mathrm{circ}}$, $\mathcal{C}_{\mathrm{Kol}}$ and $\mathcal{C}_{\mathrm{tot}}$ separately over the $(k,\theta)$ grid, Figure~\ref{fig:heatmaps} shows the two cost fields to be qualitatively different: $\mathcal{C}_{\mathrm{circ}}$ inherits the discrete plateaus of the \texttt{HT} enumeration (integer level sets in $|s|$), while $\mathcal{C}_{\mathrm{Kol}}$ varies smoothly with the symbol statistics of the winning string. 
As a consequence the $\mathcal{C}_{\mathrm{tot}}$ field is not a monotone reshading of either component alone.

\subsection{$\mathcal{C}_{\mathrm{tot}}$ optima distinctness}
\label{sec:exp-witness}

To verify our claim, we need a target at which the joint cost selects a candidate that neither axis alone would have selected. 
To detect these, the enumeration is extended to record, for each target $U$, the full list of \texttt{HT}-strings $s$ with $|s| \le M$ and $\|U_s - U\| \le \varepsilon$ rather than only the minimum-length winner. A target is flagged as a split when
\begin{equation}
  \arg\min_s \mathcal{C}_{\mathrm{tot}}(s) \;\neq\; \arg\min_s \mathcal{C}_{\mathrm{circ}}(s) \;\neq\; \arg\min_s \mathcal{C}_{\mathrm{Kol}}(s).
\end{equation}

With $\alpha=\beta=1$ on \texttt{bzip2} bits and \texttt{HT}-lengths the description term dominates by more than an order of magnitude, the joint minimizer collapses onto $\arg\min\,\mathcal{C}_{\mathrm{Kol}}$, and the split count drops to zero. 
We therefore calibrate $(\alpha,\beta)$ so the two surrogates contribute on comparable absolute scales.
A logarithmic sweep over $\beta \in [10^{-3},10]$ with $\alpha=1$ exposes a plateau $\beta \in [0.063,0.086]$ on which the split count is stable at $59$ targets. 
We adopt $(\alpha,\beta)=(1,\,0.07)$ as the representative point for downstream results. 
The split fraction is then $59/20\,327 \approx 0.3\%$; on the median split target the $\mathcal{C}_{\mathrm{tot}}$ minimizer pays a $+1$-unit penalty on $\mathcal{C}_{\mathrm{circ}}$ in exchange for an $8$-bit saving on $\mathcal{C}_{\mathrm{Kol}}$.
This exemplifies the trade-off the additive objective is designed to mediate.

\begin{figure}[htb]
  \centering
  \includegraphics[width=0.65\linewidth]{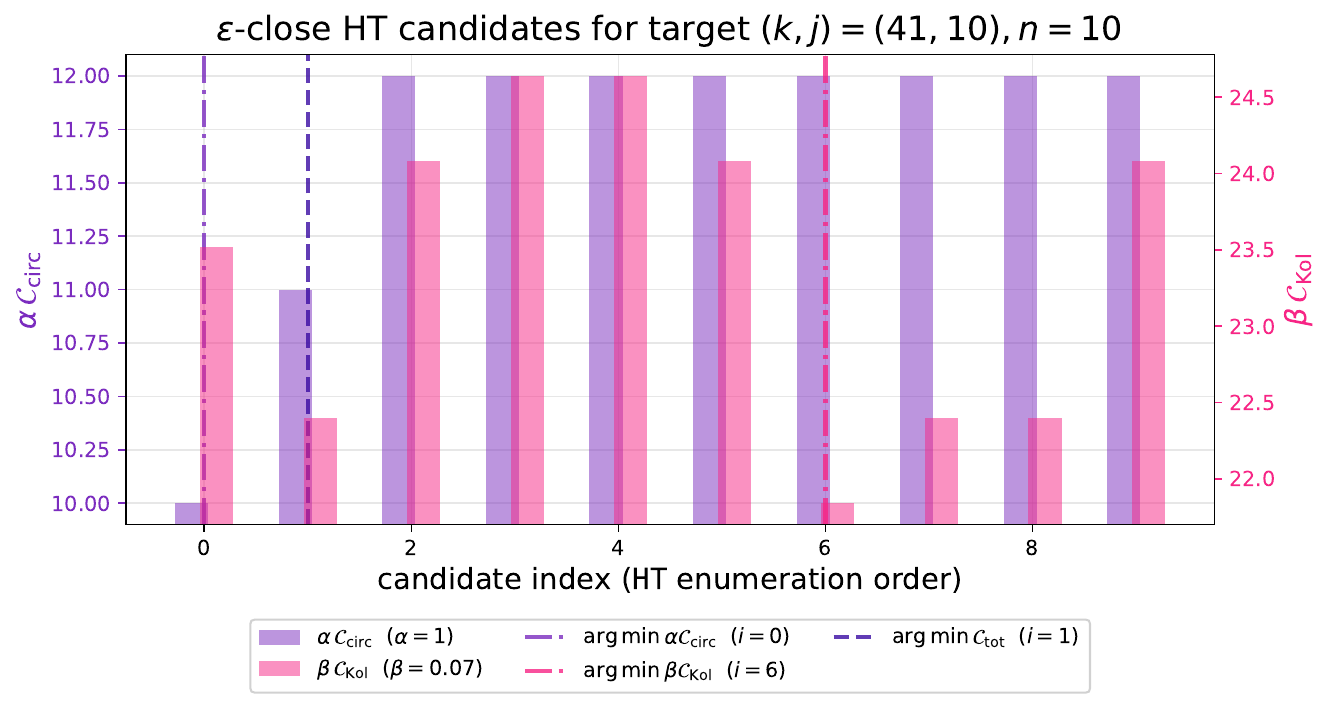}
  \caption{$\mathcal{C}_{\mathrm{tot}}$ optima distinctness for a single Fibonacci-grid target $(k,j)=(41,10)$ at $(x,y,\varepsilon,M)=(512,64,0.15,12)$ with weights $(\alpha,\beta)=(1,0.07)$. Horizontal axis: \texttt{HT}-enumeration index of the $10$ strings inside the $\varepsilon$-ball. Left axis: $\alpha\,\mathcal{C}_{\mathrm{circ}}$ (circles); right axis: $\beta\,\mathcal{C}_{\mathrm{Kol}}$ (\texttt{bzip2} bits, squares). Vertical guides mark $\arg\min\,\alpha\;\mathcal{C}_{\mathrm{circ}}$, $\arg\min\,\beta\;\mathcal{C}_{\mathrm{Kol}}$, and $\arg\min\,\mathcal{C}_{\mathrm{tot}}$; the three minimizers are at distinct decomposition candidates, so the joint cost picks a non-trivial decomposition $(11,320)$ that neither component alone selects.}
  \label{fig:tradeoff-single}
\end{figure}

Figure~\ref{fig:tradeoff-single} illustrates one such split target into the candidate-level picture. 
For the Fibonacci-grid point $(k,j)=(41,10)$, the $\varepsilon$-ball contains $10$ HT-strings. 
The per-axis minimizers $\arg\min\,\alpha\mathcal{C}_{\mathrm{circ}}$ at raw $(\mathcal{C}_{\mathrm{circ}},\mathcal{C}_{\mathrm{Kol}})=(10,336)$ and $\arg\min\,\beta\mathcal{C}_{\mathrm{Kol}}$ at $(12,312)$ each beat the other only on its own axis. 
The joint minimizer $\arg\min\,\mathcal{C}_{\mathrm{tot}}$ sits distinctly at $(11,320)$: a candidate that is neither the shortest nor the most compressible \texttt{HT}-string in the ball, but is the unique minimizer under the additive cost. 
The corresponding unitary, the output decompositions and their corresponding length, bzip encoding length and fidelity are listed below.

\begin{center}
\begin{tcolorbox}[
  colback=gray!15!black,
  colframe=gray!60,
  coltext=green!80!white,
  fontupper=\ttfamily\footnotesize,
  boxrule=0.8pt,
  arc=2pt,
  width=0.85\linewidth,
  left=6pt, right=6pt, top=6pt, bottom=6pt,
  colbacktitle=gray!35!black,
  coltitle=white,
  fonttitle=\ttfamily\small,
  title={\$ Decomposition results}
]
\begin{verbatim}
Target unitary U(k, j) = (41, 10):
  [[ 0.5582-0.6952j  0.3833+0.2412j]
   [-0.3833+0.2412j  0.5582+0.6952j]]

HT decompositions for the three minimisers:
  Ccirc (i = 0): |s| = 10, Ckol-bzip2 = 336 bits, d(U_s, U) = 0.1387 (<= 0.15)
                decomposition = T H T H T H T T T T
  Ckol  (i = 6): |s| = 12, Ckol-bzip2 = 312 bits, d(U_s, U) = 0.1387 (<= 0.15)
                decomposition = T H T H T H T H H T T T
  Ctot  (i = 1): |s| = 11, Ckol-bzip2 = 320 bits, d(U_s, U) = 0.1302 (<= 0.15)
                decomposition = T T T H T T T H T H T
\end{verbatim}
\end{tcolorbox}
\end{center}

A compiler driven by $\mathcal{C}_{\mathrm{circ}}$ alone would silently absorb the $16$-bit description-side penalty, while a compiler driven by $\mathcal{C}_{\mathrm{Kol}}$ alone (the dictionary-coding limit) would return the $(12,312)$ candidate and pay the run-time penalty. 
Only the additive objective surfaces the dominated middle option as the actual optimum.

The accompanying code repository provides an open-source implementation of this $\mathcal{C}_{\mathrm{tot}}$-guided decomposer. 
The \texttt{DeComp2.decompose} method takes a target $2\times2$ unitary and tolerance $\varepsilon$, with optional weights $(\alpha,\beta)$, a Kolmogorov surrogate (\texttt{bzip2} or \texttt{pybdm}), and a candidate backend (exhaustive \texttt{HT} enumeration, GridSynth, or Solovay-Kitaev), and returns the $\mathcal{C}_{\mathrm{tot}}$-optimal \texttt{HT} gate word, directly consumable by a quantum-compiler pipeline.

\subsection{Discussion}
\label{sec:exp-discussion}

The single-qubit study is a proof-of-principle demonstration of the compiler's combinatorial freedom. 
The following points are worth highlighting.

First, the formalism is validated only in the surrogate sense, in that $\mathcal{C}_{\mathrm{Kol}}$ is an upper bound on the true Kolmogorov complexity.
Bzip2 in particular is coarse (its output is a multiple of $8$ bits) and ignores any long-range structure beyond its window. 
The appearance of distinct minima even under such a conservative compressor therefore establishes the effect in its most conservative form.
A tighter surrogate would only enlarge the discrepancy.

Second, the split fraction of $0.3\%$ is small in absolute terms but not in operational terms.
Each split is a target on which a $\mathcal{C}_{\mathrm{circ}}$-only optimizer, asked to break a tie inside the $\varepsilon$-ball, would make the wrong choice. 
The fraction is an artifact of the discrete enumeration: \texttt{HT}-lengths are integers, so most $\varepsilon$-balls contain only one length and offer no internal trade-off. 
We expect the split fraction to rise with $M$ (denser balls), with $\varepsilon$ (more candidates per ball), with finer gate sets (e.g., a Clifford$+T$ enumerator with native $S$), and in the multi-qubit setting.

Third, the weight $\beta=0.07$ is empirical, not principled. 
It was selected as the plateau value of a sweep that maximizes the split count and therefore tells us what scale the two surrogates would need to share to be jointly informative, given the bzip2 surrogate. 
A principled $\beta$ would come from a unit-fixing argument, e.g., equating the cost of a single \texttt{HT}-symbol with the cost of one bit of description on a fixed reference machine. 

Fourth, both surrogates inherit a dependence on the gate set $\mathcal{G}$. 
The \texttt{HT} alphabet fixes the reachable lattice and hence the integer level sets that $\mathcal{C}_{\mathrm{circ}}$ inherits, while the symbol statistics that $\mathcal{C}_{\mathrm{Kol}}$ measures are defined relative to that same alphabet. 
A different universal set, whether novelty-searched as in YAQQ \cite{sarkar2026yaqq} or number-theoretically optimal super-golden gates \cite{parzanchevski2018super}, would reshape both cost fields and, in general, the location and density of the splits. 
A gate-set-independent statement of the trade-off would require descending to the pulse or continuous-control level, where the discrete alphabet dissolves into the underlying $SU(2)$ geometry, and the representation is based on the control command to an arbitrary waveform generator.

Finally, we note that the two axes are partly in tension. 
By a standard counting argument, almost all binary strings of a given length are algorithmically incompressible \cite{li2008introduction}, so a $\mathcal{C}_{\mathrm{circ}}$-optimal decomposition is typically one whose description carries near-maximal $\mathcal{C}_{\mathrm{Kol}}$. 
An incompressible minimizer admits no short generating program and is therefore invisible to any deterministic decomposition algorithm, which can only emit computable sequences. 
The additive objective is consistent with this limitation by construction, since it does not target the incompressible minimizer but the jointly cheapest reachable candidate.

\section{Future work}
\label{sec:future}

The natural next step is to lift $\mathcal{C}_{\mathrm{Kol}}$ from a passive surrogate to an active optimization signal that reorganizes the compiler IR itself.
Classical compilers have long recognized that flat instruction streams are the wrong unit of optimization. 
Three instances of this lesson translate directly to the quantum setting.
\begin{itemize}[nolistsep,leftmargin=1.2em]
  \item \textit{Macro discovery and gadget extraction.} A recurring fragment such as \texttt{H}\,\texttt{CX}\,\texttt{T} appearing $600$ times across a workload is rewritten as a single macro $M(\cdot)$, so that the optimizer manipulates $M$ instead of three primitive gates. Gadget extraction \cite{sarra2024discovering} performs exactly this on quantum circuits, and EQISA \cite{mishra2026eqisa} can package it as a controller-side mechanism.
  \item \textit{Multi-level IRs.} MLIR \cite{lattner2020mlir} deliberately retains higher-level dialects in the compilation stack rather than immediately lowering to LLVM IR, on the empirical observation that compressed, semantically richer representations are easier to optimize. QSSA \cite{peduri2022qssa} makes the analogous claim on the quantum side.
  \item \textit{Superinstructions.} Classical interpreters fuse common sequences (e.g., \texttt{LOAD}/\texttt{ADD}/\texttt{STORE} $\to$ \texttt{ADD\_STORE}) into single dispatched instructions \cite{ertl2003structure}, trading a slightly larger instruction table against shorter dispatched sequences. The same trade-off applies to a quantum controller streaming \texttt{HT}-strings, where longer per-symbol macros amortise the bit-length of each instruction over a larger circuit segment.
\end{itemize}
These reduce to a single underlying principle, that a hierarchical representation exposes repetition and parameterization, which lowers $\mathcal{C}_{\mathrm{Kol}}$ without changing the unrolled $\mathcal{C}_{\mathrm{circ}}$.

Quantum program decompilation \cite{xie2025deqompile} is a first step in this direction. 
Given a flat compiled circuit, it can recover loops, repeated templates, parameterised macros, algebraic gadgets, and a small dictionary of reusable blocks, producing a hierarchical IR whose serialisation is shorter than the input.
Its mathematical core coincides with grammar inference, dictionary learning, and minimum-description-length learning \cite{nevill1997compression,rissanen1978modeling}. 
Integrating the proposal of this article with \cite{fauquenot2025open,bach2023visualizing,mishra2026eqisa,xie2025deqompile} would address three downstream settings:
\begin{itemize}[nolistsep,leftmargin=1.2em]
  \item \textit{Quantum compiler developers.} Smaller, semantically richer IRs unlock optimization and analysis passes that are intractable on million-gate flattened streams.
  \item \textit{Quantum control and runtime engineers.} Reusable macro dictionaries reduce controller bandwidth and improve cache locality on the classical side, generalizing the EQISA compression argument to learned dictionaries.
  \item \textit{AI- and agentic-compiler pipelines.} LLM- and RL-based pass optimizers consume semantically meaningful units rather than gate-by-gate token streams, recovering the input-locality assumption these systems rely on.
\end{itemize}    
    
\section{Conclusion}
\label{sec:conclusion}

In this article, we argued that quantum compilation should optimize a life cycle cost rather than an execution-only cost.
The standard objective minimizes the circuit complexity $\mathcal{C}_{\mathrm{circ}}$, whose minimizer is highly degenerate and offers no principled way to arbitrate among the many equally short decompositions of a given target. 
We therefore paired $\mathcal{C}_{\mathrm{circ}}$ with a description-length term $\mathcal{C}_{\mathrm{Kol}}$ on the program that emits the gate sequence, yielding the additive objective $\mathcal{C}_{\mathrm{tot}} = \alpha\,\mathcal{C}_{\mathrm{circ}} + \beta\,\mathcal{C}_{\mathrm{Kol}}$ of Equation~\eqref{eq:action}. 
This serves as a minimum-description-length regularizer over the degenerate $\mathcal{C}_{\mathrm{circ}}$-minimizers, and reuses the additive execution-plus-description decomposition of the Brown-Susskind second law as a design principle.

A deterministic single-qubit experimentation is conducted on $SU(2)$ as empirical evidence of the proposed formalism. 
Exhaustive \texttt{HT} enumeration over a Haar-representative grid of targets yields a $(\mathcal{C}_{\mathrm{circ}}, \mathcal{C}_{\mathrm{Kol}})$ scatter that is visibly off-diagonal under both the \texttt{bzip2} and \texttt{pybdm} surrogates, so the description axis carries information that the circuit axis does not. 
On a small but operationally meaningful fraction of targets the joint minimizer is distinct from both per-axis minimizers, and the worked example exhibits distinctness such that the $\mathcal{C}_{\mathrm{tot}}$-minimizer is neither the shortest nor the most compressible string inside the tolerance ball.
These are the smallest non-trivial instances of the compilation choices that a $\mathcal{C}_{\mathrm{circ}}$-only optimizer discards.

Elevating $\mathcal{C}_{\mathrm{Kol}}$ from a passive analysis to an active optimization signal that drives a quantum compiler would connect $\mathcal{C}_{\mathrm{tot}}$-aware decomposition to macro discovery and quantum IR design.
The accompanying open-source implementation of the single-qubit $\mathcal{C}_{\mathrm{tot}}$ decomposer offers a starting point for such integration.

\newpage
\bibliographystyle{unsrt}
\bibliography{references}


\end{document}